\documentstyle[preprint,aps]{revtex}
\begin{document}
\draft
\preprint{CLNS 96/1392}
\title{Large $N_c$ Universality of The Baryon Isgur--Wise Form Factor: \\
The Group Theoretical Approach}
\author{Chi-Keung Chow}
\address{Newman Laboratory of Nuclear Studies, Cornell University, Ithaca, 
NY 14853.}
\date{\today}
\maketitle
\begin{abstract}
In a previous article, it has been proved under the framework of chiral 
soliton model that the same Isgur--Wise form factor describes the 
semileptonic $\Lambda_b\to\Lambda_c$ and $\Sigma^{(*)}_b\to\Sigma^{(*)}_c$ 
decays in the large $N_c$ limit.  
It is shown here that this result is in fact independent of the chiral 
soliton model and is solely the consequence of the spin-flavor SU(4) symmetry 
which arises in the baryon sector in the large $N_c$ limit.  
\end{abstract}
\narrowtext
In a previous article \cite{1}, it has been proved that the semileptonic 
$\Lambda_b\to\Lambda_c$ and $\Sigma^{(*)}_b\to\Sigma^{(*)}_c$ decays are 
controlled by the same Isgur--Wise form factor in the large $N_c$ limit.  
Recall that the semileptonic $\Lambda_b\to\Lambda_c$ decay depends on a 
universal form factor $\eta(w)$ \cite{hb1,hb2,hb3,hb4,hb5}, which is defined by
\begin{equation}
\langle \Lambda_c (v',s')|\,\bar c \Gamma b\,|\Lambda_b (v,s)\rangle = 
\eta(w)\, \bar u_{\Lambda_c}(v',s') \,\Gamma\, u_{\Lambda_b}(v,s) ,  
\end{equation}
where $w=v\cdot v'$.  
For the semileptonic $\Sigma_b^{(*)}\to\Sigma^{(*)}_c$ decay we have two 
Isgur--Wise form factors, $\zeta_1(w)$ and $\zeta_2(w)$ \cite{hb2,hb3,hb4,hb5}.
\begin{eqnarray}
&&\langle\Sigma^{(*)}_c (v',s')|\,\bar c \Gamma b\, |\Sigma^{(*)}_b(v,s)\rangle
\nonumber\\&&\; = (\zeta_1(w) g_{\mu\nu} + \zeta_2(w) v_\nu v'_\mu)\,
\bar u^\nu_{\Sigma^{(*)}_c}(v',s') \,\Gamma\, u^\mu_{\Sigma^{(*)}_b}(v,s) , 
\label{s}
\end{eqnarray}
where $u^\nu_{\Sigma^*_b}(v',s')$ is the Rarita--Schwinger spinor vector for a 
spin-$\textstyle {3\over2}$ particle and $u^\mu_{\Sigma_b}(v,s)$ is defined by
\begin{equation}
u^\mu_{\Sigma_b}(v,s) = {(\gamma^\mu + v^\mu)\gamma_5 \over \sqrt 3 }
u_{\Sigma_b}(v,s)
\end{equation}
and similarly for $u^\nu_{\Sigma^{(*)}_c}(v',s')$.  
In Ref. \cite{1}, it has been shown that 
\begin{equation}
\zeta_1(w) = -(1+w)\zeta_2 = \eta(w) .  
\label{main}
\end{equation}
i.e., the same form factor describes both semileptonic transitions.  

The large $N_c$ limit was studied in Ref. \cite{1} under the framework of 
the chiral soliton model, which is generally believed to be the realization 
of large $N_c$ QCD in the baryon sector, though the equivalence has not yet 
been rigorously proved.  
Naturally it raises the question whether the same universality of baryon 
Isgur--Wise form factor in the large $N_c$ limit can be obtained without 
reference to the chiral soliton model.  
In a recent paper \cite{2} it was attempted to get constraints on the 
Isgur--Wise form factors from unitarity in 1-pion loop renormalization of the 
$\Lambda_b\to\Lambda_c$ decay and weak decays with 1 or 2 pion emission.  
Since pion-baryon Yukawa couplings are of order $N_c^{1/2}$, many individual 
weak decay graphs involving pion lines diverge in the large $N_c$ limit.  
To preserve unitarity, non-trivial cancelation must take place between graphs 
with intermediate $\Lambda_Q$ and $\Sigma^{(*)}_Q$ states, and hence giving 
non-trivial relations between the $\Lambda_Q$ and $\Sigma^{(*)}_Q$ Isgur--Wise 
form factors.  
Ref. \cite{2} found that the relations obtained by this method are "consistent 
with, but not as powerful as" those obtained in Ref. \cite{1} through the 
chiral soliton model.  
In particular, no constraint can be placed on $\zeta_2(w)$ in Ref. \cite{2}.  
It is not surprising by noting that $\zeta_2(w)$ contributes only away from 
the point of zero recoil, where the Isgur--Wise form factors vanish like 
$\exp(-N_c^{3/2})$ in the large $N_c$ limit \cite{3}, faster than $N_c^{-n}$ 
for any finite positive $n$.  

In this article, it will be attempted to reproduce relation (\ref{main}) 
without using the chiral soliton model.  
We will use the formalism of large $N_c$ baryon developed in Ref. \cite{4}, 
which depends on group theoretical considerations and is completely model 
independent.  
It is found that relation (\ref{main}) can indeed be reproduced and hence 
the result of Ref. \cite{1} follows solely from the large $N_c$ limit 
without any additional assumptions.  

We will first review the SU(4) spin-flavor symmetry for large $N_c$ baryons
developed in Ref.~\cite{4}.  
\begin{mathletters}
The symmetry is defined by the commutation relations, 
\begin{equation}
[J^i,J^j] = i\epsilon^{ijk}J^k, \quad [I^a,I^b] = i\epsilon^{abc}I^c, \quad
[I^a,J^i] = 0, 
\label{a}
\end{equation}
\begin{equation}
[J^i, X_0^{jb}] = i\epsilon^{ijk}X_0^{kb}, \quad
[I^a, X_0^{jb}] = i\epsilon^{abc}X_0^{jc},
\label{b}
\end{equation}
\begin{equation}
[X_0^{ia}, X_0^{jb}] = 0, 
\label{c}
\end{equation}
\end{mathletters}
where $I^a$ and $J^i$ are the isospin and spin operators respectively, and 
$X_0^{ia}$ is the baryon axial current matrix element in leading order of the 
$1/N_c$ expansion, defined by
\begin{equation}
\langle B'|\bar q \gamma^i \gamma_5 \tau^a q|B\rangle 
= N_c g (X_0^{ia})_{B'B} + \hbox{higher order in $1/N_c$}.  
\label{g}
\end{equation}
Eq. (\ref{a}) is just the usual commutation relations for SU(2)$_I\,
\otimes\,$SU(2)$_J$, while Eq. (\ref{b}) states that fact that the axial 
current couplings are of spin 1 and isospin 1.  
Lastly, Eq. (\ref{c}) follows from unitary constraint of pion-baryon 
scattering and is the starting point of the formalism developed in Ref. 
\cite{4}.  

The induced representation of this spin-flavor SU(4) has been discussed in 
detail in Ref. \cite{4} and will not be repeated here.  
A vector under the induced representation is an eigenvector of $X_0^{ia}$ 
(we will denote the eigenvalue by $X_0^{ia}$ as well) and also labeled by 
its transformation properties under the little group.  
For the case of physical interest, the little group is SU(2)$\,\times Z_2$, 
and the state would be denoted as $|X_0^{ia}, K, k, \pm\rangle$ where 
$(K,k)$ and $\pm$ are the representations under the little groups SU(2) and 
$Z_2$ respectively.  
It can be shown that $\vec K=\vec I+\vec J$ and $Z_2=+$ and $-$ for 
bosonic and fermionic states respectively.  
The prime example in Ref. \cite{4} are the four nucleon states with $(I,J)=
({1\over2},{1\over2})$ and the sixteen Delta states with $(I,J)=({3\over2}, 
{3\over2})$ which fall under the {\bf 20} representation under the 
spin-flavor SU(4).  
These states can be constructed out of the basis $|X_0^{ia}, 0,0,-\rangle$, 
as $\vec K=\vec I+\vec J=0$ and both the nucleon and the Delta are 
fermions.  

To apply this formalism to heavy quark states, a straightforward application 
will be to follow the treatment of hyperons in Ref.~\cite{4} and consider the 
induced representation with $\vec K=\vec I+\vec J={1\over2}$ and $Z_2=-$.  
But it will be much more convenient and illuminating to study the induced 
representation describing just the ``brown mucks'' of the heavy baryons 
without the heavy quark.  
More exactly, instead of considering the spin-flavor SU(4) generated by 
$(I^a, J^i, X_0^{ia})$, one can consider instead that generated by $(I^a, 
s_\ell^i, X_0^{ia})$, with $s_\ell$ the spin of the ``brown muck''\footnote
{The spin of light degrees of freedom $s_\ell$ is well defined and conserved, 
a consequence of heavy quark symmetry.  
Recall that $\vec J=\vec s_Q+\vec s_\ell$.  
The conservation of $s_\ell$ follows from the conservation of the heavy quark 
spin $s_Q$ in the heavy quark limit.  
A similar treatment in the hyperon sector will be problematic as the spin of 
the strange quark is not conserved.}.  
The SU(4) commutation relations stay unchanged, and the ``brown mucks'' of 
$\Lambda_Q$ and $\Sigma^{(*)}_Q$, with $(I,s_\ell)=(0,0)$ and $(1,1)$ 
respectively, form an SU(4) {\bf 10} representation.  
Now $\vec K=\vec I+\vec s_\ell=0$ and $Z_2=+$ as the ``brown mucks'' are 
bosonic for odd $N_c$.  
Hence the heavy quark ``brown muck'' states can be constructed out of 
$|X_0^{ia}, 0,0,+\rangle$.  
Since we are not going to concern about the transformation properties under 
the little group for the rest of our discussion, these state will be 
denoted simply as $|X_0^{ia}\rangle$ below.  

It is in place to discuss the properties of the states $|X_0^{ia}\rangle$.  
Since a change of the normalization of $X_0^{ia}$ is equivalent to a 
redefinition of the axial current coupling constant $g$ in Eq.~(\ref{g}), 
we can, without loss of generality, impose the renormalization that 
\begin{equation}
X_0^{ia} X_0^{ia} = \hbox{Tr}X_0^2 = 3.  
\end{equation}
\begin{mathletters}
The states with different $X_0^{ia}$ eigenvalues can be rotated or 
iso-rotated into each other.  
\begin{equation}
U_{s_\ell}(g)|X_0^{ia}\rangle = |D^{ij}(g) X_0^{ja}\rangle, 
\end{equation}
\begin{equation}
U_I(h)|X_0^{ia}\rangle = |D^{ab}(h) X_0^{ib}\rangle, 
\end{equation}
where $U_{s_\ell}(g)$ is the unitary transformation corresponding to a finite 
spin rotation by the $g\in\hbox{SU(2)}_{s_\ell}$, $D^{ij}(g)$ is the usual 
rotation matrix in 3-dimensions, while $U_I(h)$ and $D^{ab}(h)$ are the 
counterparts in isospace for $h\in\hbox{SU(2)}_I$.  
\end{mathletters}
Moreover, since $\vec K=\vec I+\vec s_\ell=0$, for any state 
$|X_0^{ia}\rangle$ and any $g\in\hbox{SU(2)}_{s_\ell}$, there exist a certain 
$h\in\hbox{SU(2)}_I$ such that 
\begin{equation}
U_{s_\ell}(g)|X_0^{ia}\rangle = U_I(h)|X_0^{ia}\rangle, 
\label{e}
\end{equation}
i.e., an isorotation is equivalent to a rotation.  
In particular, if we choose $X_0^{ia}=\overline X_0\equiv\hbox{diag}(1,1,1)$, 
Eq.~(\ref{e}) is 
satisfied by $g=h$.  

This opens up the possibility of labeling the set of states $\{|X_0^{ia}
\rangle: \hbox{Tr} X_0^2=3\}$ by SU(2) elements.  
With the definition
\begin{equation}
|X_h\rangle = U_I(h) |\overline X_0\rangle, 
\label{h}
\end{equation}
the set $\{|X_h\rangle: h\in SU(2)_I\}$ are orthogonal.  
\begin{equation}
\langle X_{h'}|X_h \rangle = \delta(h'h^{-1}), 
\end{equation}
where $\delta(g)$ is a $\delta$-function on the SU(2) group normalized so 
that $\int dg \delta(g) = 1$.  
This association of the states to SU(2)$_I$ elements is crucial to our 
proof, as will be shown below.  

So far the heavy quark has not yet appeared in our discussion.  
In the heavy quark limit, the heavy quark is just the source of a static 
color field in which the ``brown mucks'' appear as eigenstates.  
During a $b\to c$ transition, all the ``brown muck'' feels is the change 
of the velocity of the color source.  
In this language, the Isgur--Wise form factors are just the overlap of the 
initial and final ``brown mucks'' \cite{5}.  
Now, let the state $|\overline X_0\rangle$ discussed above be one moving 
with velocity $v$.  
Of all the normalized baryon ``brown mucks'' moving with velocity $v'$, we 
will denote the one which overlaps maximally with $|\overline X_0\rangle$ as 
$|\overline X'_0\rangle$.  
Analogous to Eq.~(\ref{h}), we define 
\begin{equation}
|X'_h\rangle = U_I(h) |\overline X'_0\rangle.  
\end{equation}
Then it is trivial to prove that 
\begin{equation}
\langle X'_{h'}|\overline X_0\rangle \sim \delta(h').  
\end{equation}  
(To prove this, assume the contrary and there exist a certain non-trivial 
$h'$ for which $\langle X'_{h'}|\overline X_0\rangle> 0$.  
Then some normalized linear combination of $|\overline X'_0\rangle$ and 
$|X'_h\rangle$ will have a larger overlap with $|\overline X_0\rangle$ 
than $|\overline X'_0\rangle$, violating the assumption.)  
It follows that 
\begin{eqnarray}
\langle X'_{h'}|X_h\rangle&=&\langle X'_{h'}|U_I(h)|\overline X_0\rangle
\nonumber\\&=&\langle X'_{h'h^{-1}}|\overline X_0\rangle\sim\delta(h'h^{-1}). 
\label{o}
\end{eqnarray}

We can repeat the procedure and define a $|\overline X'_0\rangle$ for all 
$v'$.  
The overlap as a function of $w=v\cdot v'$ is denoted by $\eta(w)$ and will 
turns out to be the universal Isgur--Wise form factor.   
\begin{equation}
\langle \overline X'_0|\overline X_0\rangle = \eta(w).  
\end{equation}
A rotation in isospace gives 
\begin{equation}
\langle X'_h|X_h\rangle = \eta(w).  
\label{f}
\end{equation}
Combining Eqs. (\ref{o}) and (\ref{f}), we end up with 
\begin{equation}
\langle X'_{h'}|X_h\rangle = \eta(w) \delta(h'h^{-1}).  
\end{equation}
This is the central result of this article, the overlap of any state in 
$\{|X_h\rangle: h\in\hbox{SU(2)}\}$ with any state in $\{|X'_{h'}\rangle:
h'\in\hbox{SU(2)}\}$ can be expressed in terms of a single form factor 
$\eta(w)$.  
All remains to be done is to express the result in terms of the eigenstates 
of $I^a$ and $s_\ell^i$.  

Following the notation of Ref. \cite{1}, $|I,a;s_\ell,m\rangle$ will denote 
a state with isospin $I$ and ``brown muck'' spin $s_\ell$, while $a$ and $m$ 
are the third components of the (iso)spin.  
Then $|0,0;0,0\rangle$ and $|1,a;1,m\rangle$ are the ``brown mucks'' of 
$\Lambda_Q$ and $\Sigma^{(*)}_Q$ respectively.  
We can express $|0,0;0,0\rangle$ in terms of the $X_h$ basis.  
\begin{equation}
|0,0;0,0\rangle = \int dh |X_h\rangle, 
\end{equation}
and 
\begin{eqnarray}
\langle 0,0;0,0 (v')|0,0;0,0 (v)\rangle&=&\int dh\,dh'\,
\langle X'_{h'}|X_h\rangle\nonumber\\&=&\eta(w)\int dh\,dh'\,\delta(h'h^{-1})
\nonumber\\&=&\eta(w), 
\end{eqnarray}
justifying the notation of $\eta(w)$.  
On the other hand, 
\begin{equation}
|1,a;1,m\rangle = \int dh D^{am}(h) |X_h\rangle, 
\end{equation}
and 
\begin{eqnarray}
\langle 1,a';1,m' (v')|1,a;1,m (v)\rangle&=&\int dh\,dh'\, 
\langle X'_{h'}|{D^{a'm'}}^\dagger(h') D^{am}(h)|X_h\rangle\nonumber\\
&=&\eta(w)\int dh\,dh'\,\delta(h'h^{-1}){D^{a'm'}}^\dagger(h') D^{am}(h)
\nonumber\\&=&\eta(w) \delta_{aa'} \delta_{mm'},   
\end{eqnarray}
by the orthonormality of the rotational matrices $D^{am}$'s.  
This is exactly (the second equality of) Eq. (16) of Ref. \cite{1}.  
In terms of the full baryon states $|\Sigma^*_Q\rangle$, the result is  
\begin{eqnarray}
\langle \Sigma^*_c (v',\epsilon',s')|\bar c\Gamma b|\Sigma^*_b (v,\epsilon,s)
\rangle ={\eta(w)\over1+w}[(1+w)g_{\mu\nu}-v_\nu v'_\mu]
\epsilon'^{*\nu} \epsilon^{\mu} \bar u_c\Gamma u_b, 
\end{eqnarray}
where $\epsilon$ and $\epsilon'$ are the polarization vectors and $s$ and $s'$ 
are the heavy quark spins\footnote{As noted in Ref. \cite{2}, the ``east 
coast'' metric $(-,+,+,+)$ is used.}.  
And when compared to Eq. (\ref{s}), the main result of Ref. \cite{1} is 
recovered.  
\begin{equation}
\zeta_1(w)=-(1+w)\zeta_2(w)=\eta(w).  
\end{equation}

Since the proof above is quite complicated, let's consider an simple but 
problematic alternative proof which may help to bring out the essence of the 
correct proof above.  
Note that 
\begin{equation}
|1,a;1,m\rangle = X_0^{ma}|0,0;0,0\rangle, 
\end{equation}
and hence 
\begin{eqnarray}
\langle 1,a';1,m' (v')|1,a;1,m (v)\rangle
&=&\langle 0,0;0,0 (v)|{{X'}_0^{m'a'}}^\dagger X_0^{ma} |0,0;0,0 (v)\rangle 
\nonumber\\&=&\langle 0,0;0,0 (v')|0,0;0,0 (v)\rangle, 
\end{eqnarray}
if one sloppily identifies the operators $X_0$ and $X'_0$.  
But such sloppiness is problematic.    
The operators $X_0^{ma}$ carries a spatial index $m$, and the operator may get 
non-trivially transformed when boosted from the $v$ frame to the $v'$ frame.  
The isospin operator $I^a$, on the other hand, is manifestly independent of 
Lorentz frames.  
That is why the $|X_h\rangle$ basis is used: for these states, a rotation in 
the real space is equivalent to a rotation in the isospace.  
So, after identifying {\it one} state with velocity $v$ with {\it one} with 
velocity $v'$ (through the criterion of maximal overlap), one can establish 
a one-one correspondence between the two sets of states just by isorotations, 
which are frame-independent operations.  
These one-one corresponded states all have the same overlap, and that is the 
Isgur--Wise form factor.  

It must be emphasized that this study does {\it not} question the 
computational correctness of Ref. \cite{2}.  
While the authors of Ref. \cite{2} try to obtain constraints on the form 
factor through unitarity, the present work depends mainly on the SU(4) 
symmetry structure of baryons in the large $N_c$ limit.  
Since this SU(4) spin-flavor symmetry is completely model independent, and all 
existing approaches to large $N_c$ baryons (chiral soliton, Hartree--Fock, 
etc.) exhibit this symmetry, our result is truly model independent.  
In fact, this universality of baryon Isgur--Wise form factor should hold in 
any formalism exhibiting this spin-flavor symmetry, no matter it is large 
$N_c$ motivated or not.  
One such example is the constituent quark model developed in Ref. \cite{6}, 
which is not directly related to $1/N_c$ expansion but embodies the same 
symmetry.  
In fact, the universality of baryon Isgur--Wise form factor made its first  
appearance in their work, which chronologically precedes Ref. \cite{1}.  

In conclusion, it is found that the large $N_c$ universality of baryon 
Isgur--Wise form factor discussed in Ref. \cite{1} is solely the consequence 
of the SU(4) spin-flavor symmetry and is independent of any dynamical 
assumptions.  
This universality can be put to experimental test in the future, when more 
data are available on $\eta(w)$ (from $\Lambda_b\to\Lambda_c$ decays) and 
the $\zeta(w)$'s (from $\Omega_b\to\Omega_c$ decays).  
The deviation from universality is a measure of the (in)applicability of the 
large $N_c$ expansion for heavy baryons.  

\acknowledgements
I am grateful to Tung--Mow Yan for discussions.  
This work is supported in part by the National Science Foundation.

\end{document}